\documentclass[reprint, aps, prx, notitlepage, footinbib,longbibliography,floatfix,superscriptaddress]{revtex4-2}

\usepackage[utf8]{inputenc}
\usepackage{float}
\usepackage{graphicx}  
\usepackage{physics}
\usepackage{esint}
\usepackage{amssymb}  
\usepackage{mathtools}
\usepackage{amsmath}
\usepackage{array}
\usepackage{amsthm}
\usepackage{mathrsfs}
\usepackage[dvipsnames]{color,xcolor,colortbl}
\usepackage[colorlinks=true,linkcolor=NavyBlue,citecolor=NavyBlue,urlcolor=NavyBlue,plainpages=false,pdfpagelabels]{hyperref}
\usepackage{bm}
\usepackage[normalem]{ulem}

\newcommand{\be}{\nopagebreak[3]\begin{equation}}
\newcommand{\ee}{\end{equation}}

\newcommand{\bea}{\begin{eqnarray}}
\newcommand{\eea}{\end{eqnarray}}

\begin{document}
 
\title{Entanglement from rotating black holes in thermal baths}

\author{Ivan Agullo}
\email{agullo@lsu.edu}
\affiliation{Department of Physics and Astronomy, Louisiana State University, Baton Rouge, LA 70803, U.S.A.}
\affiliation{Perimeter Institute for Theoretical Physics, 31 Caroline Street North, Waterloo, Ontario, Canada N2L 2Y5}

\author{Anthony J. Brady}
\email{ajbrad4123@gmail.com}
\affiliation{Department of Electrical and Computer Engineering, University of Arizona,  Tucson, Arizona 85721, USA}

\author{Adri\`a Delhom}
\email{adria.delhom@gmail.com}
\affiliation{Department of Physics and Astronomy, Louisiana State University, Baton Rouge, LA 70803, U.S.A.}
\affiliation{Laboratory of Theoretical Physics, Institute of Physics, University of Tartu, W. Ostwaldi
1, 50411 Tartu, Estonia}

\author{Dimitrios Kranas}
\email{dkrana1@lsu.edu}
\affiliation{Department of Physics and Astronomy, Louisiana State University, Baton Rouge, LA 70803, U.S.A.}

\begin{abstract}

We extend previous efforts to quantify the entanglement generated in Hawking’s evaporation process by including rotation and thermal environments (e.g. the cosmic microwave background). Both extensions are needed to describe real black holes in our universe. Leveraging techniques from Gaussian quantum information, we find that the black hole's ergoregion is an active source of quantum entanglement and that thermal environments drastically degrade entanglement generation. Our predictions are suitable to be tested in the lab using analogue platforms and also provide tools to assess the fate of quantum information for black holes in more generic settings. \\
\end{abstract}

\maketitle

\section{Introduction}  In 1974, Stephen Hawking showed that black holes emit radiation as hot bodies \cite{Hawking74,Hawking:1975vcx}, providing an unforeseen relation between relativity, thermodynamics, and quantum mechanics.  
The quantum nature of this process becomes manifest by noticing that the emitted  Hawking radiation is {\em entangled} with modes that fall into the interior of the black hole.

A milestone in the characterization of entanglement generated by non-rotating black holes was achieved by D. Page in Refs.~\cite{PageCurve93PRL,page2013JCAP} (see also Ref.~\cite{zurek1982PRL}).  
Page considered  quantum fields initially prepared in the vacuum state. Since this is a pure quantum state,  the von Neumann entropy of the radiation emitted to infinity quantifies the entanglement with the radiation falling inside the black hole. These calculations generated great interest; specifically because they provide the tools to discuss the information loss paradox \cite{Hawking:1976ra,UnruhWald17, marolf2017BHinfoRvw,ashtekar2020BHevapLQG,almheiri2021RMP} in quantitative terms \cite{PageCurve93PRL,page2013JCAP}.  

The goal of this article is to extend Page's calculations in two important directions: (i) incorporating rotation and (ii) accounting for thermal environments. Both generalizations are essential for a comprehensive description of realistic black holes in our universe. For instance, rotation---prevalent for black holes of astrophysical origin---gives rise to an ergoregion encompassing the black hole horizon. It is well understood that the ergoregion contributes to Hawking radiation by amplifying its intensity via the phenomenon of superradiance~\cite{Hawking:1975vcx,Wald:1975kc,starobinskii1973superrad,unruh1974superrad}.

In this article, we elucidate that entanglement generated by a rotating black hole contains two contributions of different physical origin---a thermal contribution from the horizon and a non-thermal part from the ergoregion. Both contributions are described mathematically in similar terms, by two-mode squeezers.  A notable distinction is that the ergoregion is seeded by the thermal Hawking radiation from the horizon. 
The total entanglement results from the interplay of these two phenomena. To the best of our knowledge, there has been no previous investigation into characterizing the entanglement generated by the ergoregion.

That the evolution of field modes across the ergoregion is formally described by a two-mode squeezer was first noticed in \cite{davies1993QuantumSuperradiance}. This implies that the vacuum state undergoes a transformation into a  two-mode squeezed vacuum made of entangled pairs of particles—this is a vacuum instability, first pointed out by Starobinski \cite{starobinskii1973superrad} and confirmed by Unruh \cite{unruh1974superrad} (see also \cite{AshtekarMagnon1975} for a discussion on a related instability of rotating compact stars that feature an ergoregion.).

The scenario considered in this article differs from the Starobinski-Unruh framework in that, if the black hole originates from gravitational collapse, quantum fields in the vicinity of the ergoregion are not in the vacuum state, but instead contain a thermal flux of particles (Hawking radiation) emanating from the outer horizon.  The ergoregion amplifies these thermal Hawking quanta by the process of superradiance, 
 imprinting additional entanglement in the radiation. This contribution to entanglement differs from the one that the ergoregion would create if Hawking radiation were not present, as in the idealized Starobinski-Unruh scenario. Although the total radiation emitted by rotating black holes has been extensively studied, including the combined contribution of the horizon and the ergoregion \cite{Hawking:1975vcx,Wald:1975kc,Page:1976ki}, no analysis of entanglement---further decoupling each contribution---has been explored so far.

We furthermore demonstrate that, for black holes immersed in thermal baths, the exterior thermal radiation alters the entanglement generated by the black hole, appreciably degrading entanglement if the temperature of the environment is larger than the Hawking temperature. This fact has been recently  pointed out by us in the context of analogue horizons created in the lab \cite{ brady2022SympCircs,agullo2022prl} (and qualitatively discussed for non-rotating black holes in the essay~\cite{Essay}), and is of obvious physical interest for astrophysical black holes, considering that all of them are immersed in the cosmic background radiation. Furthermore, thermal environments are ubiquitous in analogue black holes created in laboratory settings.

The quantitative analysis of these effects requires other tools than the ones used by Page. In particular,  thermal environments are described by mixed quantum states and, consequently, the von Neumann entropy of the outgoing radiation is no longer a quantifier of entanglement. We use a novel way of deriving the Hawking process based on general techniques used to describe Gaussian bosonic quantum systems. These tools make it possible to derive the Hawking effect in a remarkably simple yet powerful manner, allowing us to isolate and quantify the contribution of the ergoregion.

On the other hand, our tools  only apply to states of the fields that are Gaussian (either pure or mixed). Although this is an intrinsic limitation, the formalism includes many of the states that are physically interesting: vacuum, coherent, squeezed and thermal states. Hence,  the limitation is not significant in practice.

In what follows, we neglect contributions from particle species with rest-mass energies greater than the thermal energy of Hawking radiation, which, for solar-mass black holes or heavier, practically restricts our calculations to the known massless particles, namely photons and gravitons~\cite{page2013JCAP}. We use units in which $\hbar=c=k_B=1$. 

\section{Modes and their evolution}  
\begin{figure}
{\centering   
    \includegraphics[width=.75\linewidth]{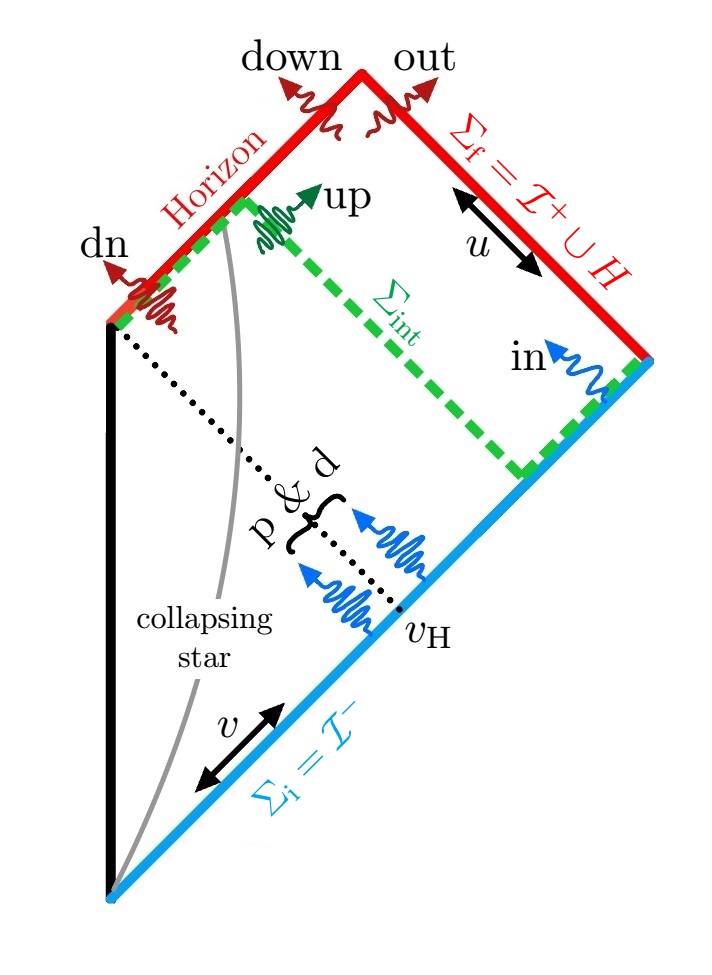}
      \caption{Spacetime representation of the Hawking process (only the region exterior to the outer horizon is shown). For each choice of $\omega$, $\ell$, $m$, $s$, the Hawking effect results from the evolution of three modes at the  Cauchy hypersurface $\Sigma_{\rm i}$ to three modes at the  Cauchy hypersurface $\Sigma_{\rm f}$. The intermediary mode ``up'' is introduced to describe the evolution through the  Cauchy hypersurface $\Sigma_{\rm int}$ and to separate the total evolution in two steps.
      }
     \label{fig:Plot1}
  }
\end{figure}

Figure~\ref{fig:Plot1} shows the Penrose diagram of  the formation of a rotating, axisymmetric black hole (only the region exterior to the outer horizon is shown). Following Hawking's original derivation, the evolution of quantum fields can be obtained by solving the classical wave scattering. Let us first identify the wave modes, describing either electromagnetic or gravitational waves, involved in the Hawking effect (see Fig.~\ref{fig:Plot1}). 

Consider an ``out'' mode at future null infinity ($\mathcal{I}^+$). This is described by a wave packet sharply centered around a frequency $\omega$, angular momentum numbers $\ell$ and $m$, and helicity $s$, and  supported mostly at late values of the retarded time $u$ along $\mathcal{I}^+$. We will denote this mode by $\varphi^{(\rm out)}_Q$, where $Q=\omega, \ell, m, s$, and we have suppressed vectorial and tensorial indices in the modes. The evolution of this mode backward in time to past null infinity  ($\mathcal{I}^-$) produces a combination of three modes, which, following \cite{Frolov:1998wf}, we denote as $\varphi^{(\rm in)}_Q$, $\varphi^{(\rm p)}_Q$ and $\varphi^{(\rm d)}_Q$. The mode $\varphi^{(\rm in)}_Q$ is a wave packet with support centered at late values of the advanced time $v$ in $\mathcal{I}^-$, in the region where the spacetime is  described by Kerr's metric. Due to  the stationary and axisymmetric character of the underlying geometry in this region of the spacetime, $\varphi^{(\rm in)}_Q$ has the same frequency and angular momentum numbers as $\varphi^{(\rm out)}_Q$.

The modes $\varphi^{(\rm p)}_Q$ and $\varphi^{(\rm d)}_Q$ have support only around the time $v_H$ at which the horizon forms. Although they have the same $\ell$ and $m$ numbers as  $\varphi^{(\rm out)}_Q$, they do not have a well-defined frequency at $\mathcal{I}^-$. Due to the blueshift induced by the horizon, these modes have support on exponentially large frequencies. Hence, the label $\omega$ of these modes should not be confused with their physical frequency at $\mathcal{I}^-$.
The explicit expression of these modes can be found in  Appendix \ref{sec:2}. These  modes in $\mathcal{I}^-$, which describe the progenitors of the  ``out'' mode, were introduced in \cite{Wald:1975kc}, and are known as Wald basis modes (see \cite{Frolov:1998wf} for a detailed account). The use of Wald basis introduces a drastic simplification---factorizing the evolution of infinitely many modes into decoupled $Q$-sectors and reducing the analysis to evolving three modes at a time.

If we start with an arbitrary combination of  $\varphi^{(\rm p)}_Q$, $\varphi^{(\rm d)}_Q$ and  $\varphi^{(\rm in)}_Q$ at $\Sigma_{\rm i}=\mathcal{I}^-$,  the evolution produces, at the final Cauchy hypersurface $\Sigma_{\rm f}=\mathcal{I}^+\cup H$ ($H$ denotes the event horizon), a combination of three modes, $\varphi^{(\rm out)}_Q$ and two extra modes which we denote by $\varphi^{(\rm dn)}_Q$ and $\varphi^{(\rm down)}_Q$ (see Fig.~\ref{fig:Plot1}). The evolution is made of two contributions of distinct physical origin, which we discuss separately. The first contribution corresponds to the evolution from $\Sigma_{\rm i}$ to $\Sigma_{\rm int}$ (the form of the Cauchy hypersurface  $\Sigma_{\rm int}$ is depicted in Fig.~\ref{fig:Plot1}; ``int'' stands for ``intermediate''). This part of the evolution occurs in the region of spacetime that is time-dependent, in which the horizon is formed. The evolution from $\Sigma_{\rm int}$ to $\Sigma_{\rm f}$ takes place in the stationary Kerr geometry.

For the free field theories we consider in this paper, the evolution is linear and can be written as matrix multiplication as follows. Let 
\be \vec{A}_Q^{(\rm i)}\equiv (a^{(\rm in)}_Q,\,  a^{(\rm p)}_Q,\, a^{(\rm d)}_Q, a^{(\rm in)\, \dagger}_Q,\,  a^{(\rm p)\, \dagger}_Q,\,  a^{(\rm d)\, \dagger}_Q)^{\top}\,  \ee
be the column vector of annihilation and creation operator associated with the modes defined in $\Sigma_{\rm i}$, and let
\be \vec{A}_Q^{(\rm int)}\equiv(a^{(\rm in)}_Q,\,  a^{(\rm up)}_Q,\, a^{(\rm dn)}_Q, a^{(\rm in)\, \dagger}_Q,\,   a^{(\rm up)\, \dagger}_Q,\,  a^{(\rm dn)\, \dagger}_Q)^{\top}  \ee
 be a similarly defined vector associated with the modes at ${\Sigma_{\rm int}}$. The modes ``dn'' and ``up'' are depicted in Fig.~\ref{fig:Plot1}, and their explicit form is written in the Appendix \ref{modes}. Let us consider a black hole with mass $M$ and angular momentum $J$. The surface gravity is  $\kappa= (r_+-r_-)/[2(r_+^2+a^2)]$, where $r_{\pm}=M\pm\sqrt{M^2-a^2}$,   $a\equiv J/M$ and $\Omega_H= a\, {(r_+^2+a^2)}^{-1}$ is the ``angular velocity'' of the horizon.

The Heisenberg evolution from $\Sigma_{\rm i}$ to $\Sigma_{\rm int}$ is encoded in the relation  $\vec{A}_Q^{(\rm int)}={\bf S}^{(\rm H)}_Q\cdot \vec{A}_Q^{(\rm i)}$, where 
${\bf S}^{(\rm H)}_Q=\begin{pmatrix}{\bf A}_Q & {\bf B}_Q  \\ 
  {{\bf B}}_Q  & {{\bf A}}_Q 
    \end{pmatrix}$ and
 \begin{align}
 {\bf A}_Q&=\begin{pmatrix}1 &0  \\  0 & \cosh z_H \, \mathbb{I}_2   \end{pmatrix},\\
 \ {\bf B}_Q&=\begin{pmatrix} 0 &0 \\  0 & \sinh z_H \, \bar{\mathbb{I}}_2 \end{pmatrix},
  \end{align}
The label H in  ${\bf S}^{(\rm H)}$ stands for horizon. Here, $z_H(\omega,m)= \tanh^{-1} e^{-{2\pi}|\tilde \omega|/(2\kappa)}$, and $\tilde \omega\equiv \omega-m\, \Omega_H$. These expressions are valid for both photons and gravitons. We have denoted by $\mathbb{I}_n$ and $\bar{\mathbb{I}}_n$ the diagonal and anti-diagonal unit matrices in $n$ dimensions, respectively. We note that $\cosh z_H\to \infty$ as $\tilde \omega\rightarrow 0$; however, this mathematical divergence is tamed by the evolution across the ergoregion because the transmission probability across the ergoregion vanishes in the limit $\tilde \omega \to 0$.  The elements of ${\bf S}^{(\rm H)}_Q$, as well as those of other evolution matrices written below, are called Bogoluibov coefficients. These coefficients are derived by solving the classical equations of motion, as detailed in the Appendix \ref{sec:2}.

Interestingly,  the form of ${\bf S}^{(\rm H)}_Q$ reveals that this operation precisely corresponds to a {\em two-mode squeezing process} between  modes ``p'' and ``d'', to produce ``dn'' and ``up'' (and the identity operation for ``in'').  A two-mode squeezer converts the vacuum state into a ``two-mode squeezed vacuum'', which is an entangled Gaussian state such that the reduced state of each individual mode is a mixed thermal state. The occupation number of either mode is equal to $\bar n=\sinh^2 z_H=(e^{2\pi |\tilde \omega|/\kappa}-1)^{-1}$, corresponding to a black body spectrum at temperature $\kappa/2\pi$. 

The evolution from $\Sigma_{\rm int}$ to $\Sigma_{\rm f}$ describes modes propagating across the ergoregion encompassing the horizon. This evolution is starkly different for modes with $\tilde \omega>0$ (non-superradiant modes, NSRM) versus modes with $\tilde \omega<0$ (superradiant modes, SRM). For NSRM, we have $\vec{A}_Q^{{\rm (f)}}={\bf S}^{(\rm NSR)}_{Q}\cdot \vec{A}_Q^{\rm(int)}$, where
\be   \vec{A}_Q^{\rm (f)}\equiv ( a^{(\rm out)}_Q, \, a^{(\rm down)}_Q,a^{(\rm dn)}_Q , a^{(\rm out)}_Q{}^\dagger,\, a^{(\rm down)}_Q{}^\dagger,\, a^{(\rm dn)}_Q{}^\dagger)^\top\, \ee and ${\bf S}^{(\rm NSR)}_Q=\begin{pmatrix}{\bf C}_Q & 0  \\ 
 0  & {\bf C}_Q 
    \end{pmatrix}$, with
 \be \label{BS} {\bf C}_Q=\begin{pmatrix} \sin \theta_Q& \cos \theta_Q & 0 \\  \cos \theta_Q & - \sin \theta_Q & 0 \\ 0& 0 & 1  \end{pmatrix} .\ee 
The term $\cos^2 \theta_Q$ represents the transmission coefficient of the potential barrier (i.e., the ``greybody'' factor) and depends on $\omega$, $\ell$, $m$ and the spin $s$. Hence, these coefficients differ for electromagnetic and gravitational waves.

 The transformation \eqref{BS} corresponds to a {\em beam splitter}. Beam splitters are {\em passive  transformations}, meaning they do not amplify waves (no particle creation).  Their role is simply to divide the incoming waves in transmitted and reflected portions.

For SRM, the evolution across the ergoregion has a {\em completely different character}, 
and is described by 
 ${\bf S}^{(\rm SR)}_{Q}=\begin{pmatrix}{\bf A}^{\rm SR}_Q & {\bf B}^{\rm SR}_Q  \\ 
 {\bf B}^{\rm SR}_Q  & {\bf A}^{\rm SR}_Q 
    \end{pmatrix}$
with 
\begin{align}
{\bf A}^{\rm SR}_Q&=\begin{pmatrix}\cosh z_{\rm erg} \, \mathbb{I}_2  &0  \\  0 & 1   \end{pmatrix}, \\
{\bf B}^{\rm SR}_Q&= \begin{pmatrix}\sinh z_{\rm erg} \, \bar{\mathbb{I}}_2 & 0  \\  0 & 0 \end{pmatrix} .
\end{align}
This is a two-mode squeezing process between modes ``in'' and ``up'' to produce ``out'' and ``down'' \cite{davies1993QuantumSuperradiance}. The squeezing intensity $z_{\rm erg}$ (``erg'' stands for ergoregion) depends on $\omega$, $\ell$, $m$ and $s$. 
This implies that, for SRM, the ergoregion acts like a quantum amplifier---able to create entangled pairs of particles---very much like the horizon does. Note that the evolution matrices are real; this is possible by exploiting the freedom to introduce a time-independent phase to the aforementioned creation and annihilation operators.


The transition of the ergoregion---from acting like a beam splitter for NSRM to a two-mode squeezer for SRM---is of vital importance for most results presented in this article. From a physical perspective, this transition arises from the fact that the Killing field describing time translations at infinity becomes spacelike within the ergoregion. Consequently, the modes labeled ``down" having $\tilde \omega <0$ (although $\omega>0$) carry negative energy across the horizon from the standpoint of inertial observers at infinity. The appearance of these negative energy modes within the ergoregion provides a mechanism for energy extraction, effectively transforming the ergoregion into an amplifier. This line of reasoning builds upon the ideas of Penrose and Zeldovich~\cite{Penrose:1969pc,Penrose:1971uk,Zeldovich1,Zeldovich2}. Our goal is to extend their implications into the quantum realm by demonstrating that this mechanism for energy extraction also acts as a source of entanglement.

The complete evolution,  from $\Sigma_{\rm i}$ to $\Sigma_{\rm f}$, is described by the matrix  ${\bf S}^{(\rm tot)}_Q$, obtained by simply multiplying the evolution matrices of the two processes. For NSRM, ${\bf S}^{(\rm tot)}_Q={\bf S}^{(\rm NSR)}_Q\cdot{\bf S}^{(\rm H)}_Q$; while for SRM, ${\bf S}^{(\rm tot)}_Q={\bf S}^{(\rm SR)}_Q\cdot{\bf S}^{(\rm H)}_Q$. Given a quantum state for electromagnetic and gravitational perturbations at $\Sigma_{\rm i}$, ${\bf S}^{(\rm tot)}$ is all we need to compute observable quantities at $\Sigma_{\rm f}$. We have calculated both $\theta_Q$ and $z_{\rm erg}$ (thereby characterizing $\bm S_{Q}^{(\rm tot)}$) by numerically solving Teukolsky's equation in the Kerr geometry for electromagnetic and gravitational waves, following Teukolsky-Press \cite{Teukolsky:1974yv} and Page \cite{Page:1976df,Page:1976ki}. Some details of our numerical calculations can be found in the Appendix \ref{num}. This is the only place where numerics enters in our analysis (which includes the computation of the angular eigenvalues). The rest, including calculations of entanglement, are obtained from analytic formulas.

\section{Gaussian state evolution.} 
The advantage of restricting to Gaussian states is that they can be described using vectors and matrices in the classical phase space, which are significantly more amenable than density matrices in the (infinite dimensional) Hilbert space. The reason is that Gaussian states are uniquely determined by their first and second moments, since higher order moments can all be derived from them. In field theory, the phase space is infinite dimensional. Though, for our problem, the evolution factorizes in decoupled $Q$-sectors, reducing the analysis to an effective six-dimensional phase space. 

Let us fix a value of $Q$. Given a Gaussian state (pure or mixed) at  $\Sigma_{\rm i}$, the information in its density matrix $\rho_Q$ is encoded in the vector of first moments $\vec \mu^{(\rm i)}_Q\equiv {\rm Tr}[\rho_Q  \,  \vec A_Q^{(\rm i)}]$ and the covariance matrix ${\bf \sigma}^{(\rm i)}_Q\equiv {\rm Tr}[\rho_Q\,  \{  \vec A_Q^{(\rm i)}-\vec \mu^{(\rm i)}_Q, \vec A_Q^{(\rm i)}-\vec \mu^{(\rm i)}_Q\}]$, where the curly brackets denote anti-commutators. The covariance matrix contains the symmetric part of the second moments, after subtracting the first moments; the anti-symmetric part is state independent and fully determined by the canonical commutation relations. For Gaussian states, the pair $(\vec \mu^{(\rm i)}_Q, {\bf \sigma}^{(\rm i)}_Q)$ completely specifies the density matrix $\rho_Q$.
Linear evolution preserves the Gaussian character of the state and, at $\Sigma_{\rm f}$, the  state is a Gaussian state fully described by $\vec \mu^{(\rm f)}_Q={\bf S}^{(\rm tot)}_Q\cdot \vec \mu^{(\rm i)}_Q$ and  ${\bf \sigma}^{(\rm f)}_Q={\bf S}^{(\rm tot)}_Q\cdot {\bf \sigma}^{(\rm i)}_Q\cdot {\bf S}^{(\rm tot)\, \top}_Q$. From the pair $(\vec \mu^{(\rm f)}_Q, {\bf \sigma}^{(\rm f)}_Q)$ we can obtain particle number, entropies, entanglement, etc. in a simple manner. These techniques are well known (see, e.g., \cite{weedbrook2012,serafini17QCV}) but a summary is included in the Appendix \ref{Gauss}. 

\begin{figure}
{\centering   
    \includegraphics[width=\linewidth]{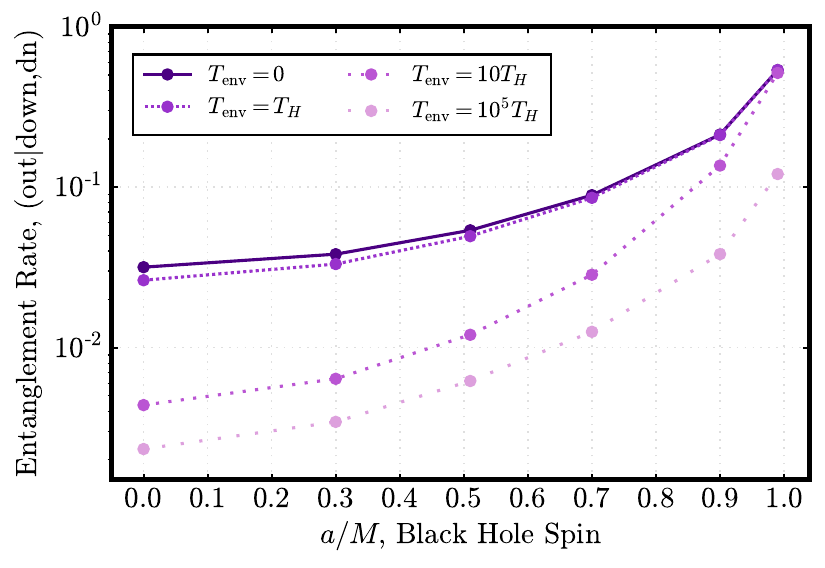}
      \caption{Entanglement rate for the partition separating the interior and exterior modes of the black hole. $T_H\equiv \kappa/(2\pi)$ is the Hawking temperature.}
     \label{fig:LogNegFlux}
  }
\end{figure}

For the initial state, we consider populating the ``in'' mode with thermal quanta. The ``p'' and ``d'' are set in vacuum because they are supported on ultra-high frequencies and, for temperatures well below the Planck scale,
it is an excellent approximation to consider them unpopulated. The initial state is mixed and described by $\vec \mu^{(\rm i)}_Q=\vec 0$ and ${\bf \sigma}^{(\rm i)}_Q=\begin{pmatrix} 0 & D_Q \\ D_Q& 0 \end{pmatrix}$, where
\be  D_Q=\begin{pmatrix}(1+2\, n^{\rm env}_{\omega})  & 0 \\ 0 &  \mathbb{I}_2\end{pmatrix}\, , \ee
and $n^{\rm env}_{\omega}=(e^{\omega/T_{\rm env}}-1)^{-1}$ is the mean number of quanta in the mode ``in'' for  environment temperature $T_{\rm env}$.

\begin{figure*}
\centering   
    \includegraphics[width=.49\linewidth]{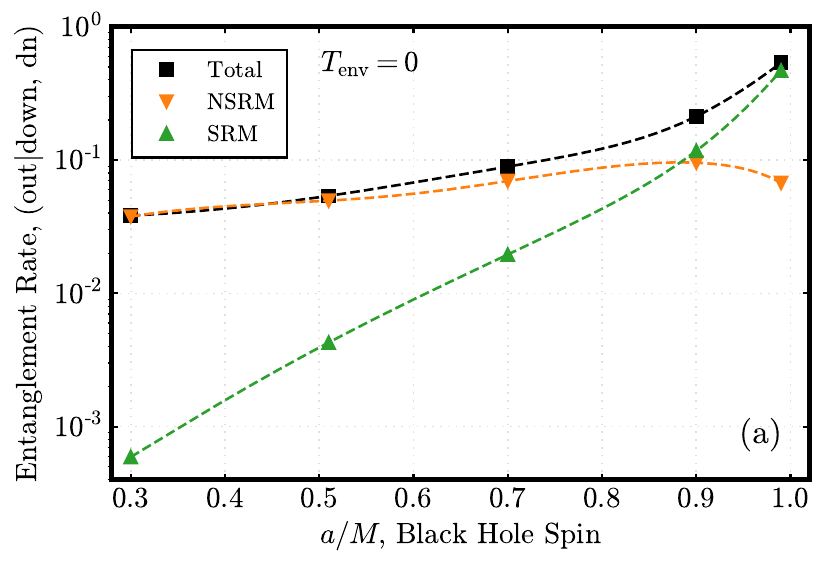}
    \includegraphics[width=.49\linewidth]{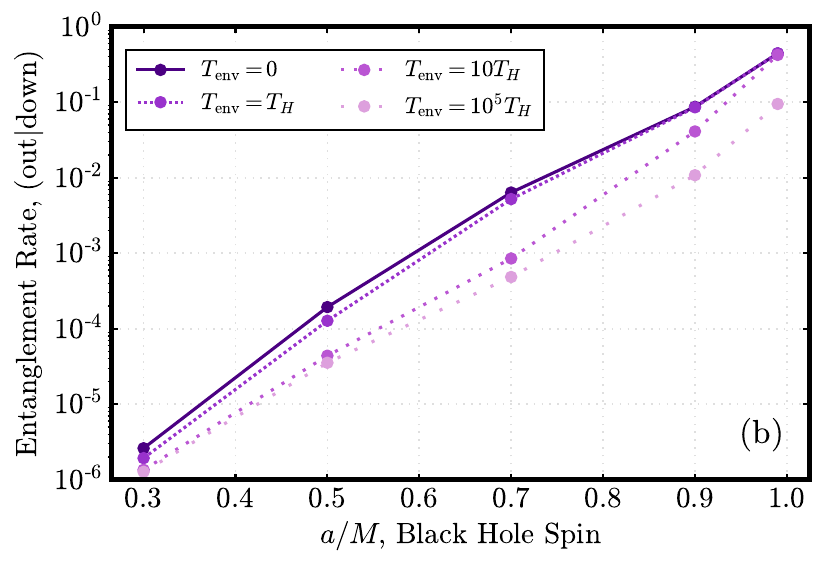}
      \caption{(a) Entanglement rate  (quantified by LN per unit time) for the partition between the interior and exterior of the black hole, separated into SRM and NSRM contributions for $T_{\rm env}=0$. (b) Entanglement rate for out$|$down modes (a ``superradiant pair'') for various $T_{\rm env}$.}
     \label{fig:SRVSNSR}
\end{figure*}

{\bf Particle emission.} It is straightforward  to read from  ${\bf \sigma}^{(\rm f)}_Q$ the mean particle number on each of the modes. We discuss NSRM and SRM separately. For NSRM modes, the number of ``out'' quanta is encoded in the component $(1,4)$ of the matrix  ${\bf \sigma}^{(\rm f)}_Q$ and given as
\be \label{nH} \bar n^{(\rm out)}_Q =\sin^2\theta_Q\, n^{\rm env}_{\omega}+\, \cos^2\theta_Q\, \sinh^2 z_H\, . \ee
The second term accounts for the Hawking quanta, where $\sinh^2 z_H$ is the thermal part of the spectrum and  $\cos^2\theta_Q$ is the greybody factor. The first term corresponds to the portion of the  thermal radiation approaching the black hole from outside that is reflected back to $\mathcal{I}^+$.

For SRM, expression \eqref{nH} is replaced by 
\be\label{nHSRM}  \bar n^{(\rm out)}_Q =n^{\rm env}_{\omega}+\sinh z_{\rm erg}^2\, (1+n^{\rm env}_{\omega}+\sinh z_H^2) . \ee
The first term corresponds to the quanta already present in the initial state, while the second terms is proportional to the quanta generated by the ergoregion-squeezer, $\sinh z_{\rm erg}^2$.  The last contribution describes the interplay of the two squeezing processes. We find that NSRM dominate the emission for $a\lesssim 0.9 M$.

\section{Entropy and Entanglement generation.}
The von Neumann entropy of the ``out'' modes can easily be obtained from the covariance matrix ${\bf \sigma}^{(\rm f)}_Q$ (see Appendix \ref{sec:3}) 
\be \label{SH} S_Q^{(\rm out)}=\big(\bar n^{(\rm out)}_Q+1\big) \ln (\bar n^{(\rm out)}_Q+1) -\, \bar n^{(\rm out)}_Q\, \ln{\bar n^{(\rm out)}_Q}.\ee
If we restrict to $a=0$ and to vacuum input ($T_{\rm env}=0$), Eqn.~\eqref{SH} reproduces Page's results \cite{page2013JCAP}. For SRM, the entropy of the Hawking radiation is again given by expression \eqref{SH}, using now Eqn.~\eqref{nHSRM} for $\bar n^{(\rm out)}_Q$. These expressions extend Page's entropy calculations~\cite{page2013JCAP} to the rotating case and for $T_{\rm env}\neq 0$. 

For mixed quantum states, the entropy of the outgoing radiation is no longer adequate for quantifying entanglement. Instead, we employ the Logarithmic Negativity~\cite{vidal02,plenio05,werner01bound}, since it is a measure of entanglement that is applicable to both pure and mixed states. Let $\rho$ be the density matrix of a quantum system made of two subsystems, A and B. The Logarithmic Negativity with respect to the partition $A|B$ is defined as
\be {\rm LN}(\rho)=\log_2 || \rho^{\top_B}||_1\, \label{eq:defLN},\ee
where $\rho^{\top_B}$ represents the partial transpose of $\rho$ with respect to subsystem  B, and $||\cdot||_1$ denotes the trace norm. A non-zero LN value indicates a violation of the Positivity of the Partial Transpose criterion for quantum states \cite{peres96}. For Gaussian states, and when one subsystem contains a single mode, LN is non-zero if and only if the state is entangled. Furthermore, LN is as a faithful quantifier of entanglement, signifying that a higher LN value corresponds to a greater degree of entanglement.

From the covariance matrix ${\bf \sigma}^{(\rm f)}_Q$, it is a simple task to compute the Logarithmic Negativity between any bi-partition $A|B$ of the three modes (see Appendix \ref{sec:3}). An interesting partition is the interior-exterior one, where subsystem A is made of the ``out'' mode and subsystem B of the two modes falling in the horizon (``dn" and ``down"). Figure~\ref{fig:LogNegFlux} shows the entanglement rate (measured via the Logarithmic Negativity generated per unit of retarded time $u$). This plot communicates two important messages: (i) rotation enhances the entanglement of the black hole with the exterior; (ii) ambient thermal radiation inhibits the generation of entanglement, reducing entanglement by $\sim 90\%$. Intuitively, this is because thermal fluctuations can overwhelm quantum coherence within the system, leading to reduced entanglement. The degradation of entanglement, in turn, illustrates that the entropy of the Hawking mode is very different from entanglement: entropy grows with the environment temperature, while entanglement decreases. It is also interesting to note that, when $T_{\rm env}\gg T_H$, even when the black hole does not evaporate, it keeps creating entanglement, although at a much slower rate than that in isolation. 


Next, we want to understand how much of the entanglement is contained in 
SRM and NSRM. The answer is revealed in Fig.~\ref{fig:SRVSNSR}. Figure~\ref{fig:SRVSNSR}(a) shows the entanglement rate for the interior-exterior partition, for $T_{\rm env}=0$, together with the contribution from SRM and NSRM. We observe that, as for energy and angular momentum, NSRM dominates for black holes with $a\lesssim 0.9M$. In other words, for $a\lesssim 0.9 M$, most of the entanglement generated can be attributed to the horizon.  

\begin{figure}[!htb]
    \centering
    \includegraphics[width=\linewidth]{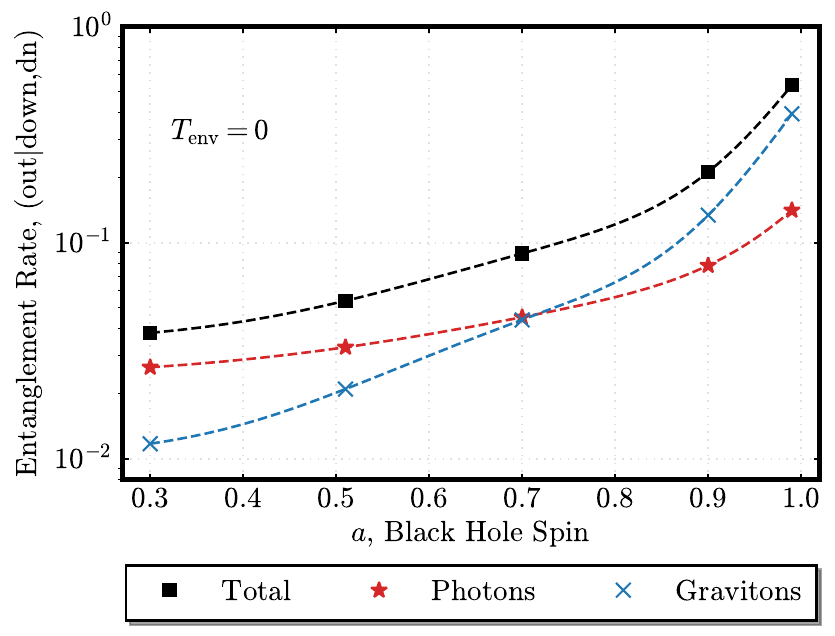}
    \caption{Entanglement rate (quantified by LN per unit time) for the partition between the interior and exterior of the black hole at $T_{\rm env}=0$. Similar behavior holds for $T_{\rm env}>0$, albeit with parametrically suppressed values of LN. We show separately the contribution from photons and gravitons.}
    \label{fig:enter-label}
\end{figure}

Finally, we want to know what portion of the entanglement in SRM can be attributed to the ergoregion squeezer. This is not obvious \textit{a priori} because SRM experience both squeezers. For this, we compute the entanglement between the ``out'' mode at $\mathcal{I}^+$ and the ``down'' mode at $H$. This entanglement  cannot be generated by the horizon-squeezer; hence, it is a measure of the contribution of the ergoregion to the final entanglement. 
Figure~\ref{fig:SRVSNSR}(b) shows that entanglement between ``out'' and ``down'' modes is very small for $a\lesssim 0.9 M$. On the contrary, for  $a> 0.9M$, the entanglement between the ``out'' and ``down'' pair is of the same order as the total entanglement between exterior and interior of the black hole, confirming that the ergoregion plays a significant role in generating entanglement. Notice that the previous plots report entanglement   carried out by both photons and gravitons together. We find that gravitons dominate the generation of entanglement for $a>0.7M$ (see Fig.~\ref{fig:enter-label}). This aligns with the findings on energy emission in \cite{Page:1976ki}. 


\section{Discussion} When the initial state of the quantum fields is assumed to be the vacuum, it suffices to look at the outgoing Hawking radiation to understand the entanglement generated by an evaporating black hole. For the goals of this article, however, a detailed characterization of the final state of modes that fall inside the horizon---and their individual correlation with Hawking modes---is desired. The Gaussian formalism employed here provides a remarkably good balance between simplicity and power, from which we have been able to quantify the impact of ambient thermal radiation and the contribution of the ergoregion to the process of Hawking radiation and entanglement generation.

The input to the horizon squeezer is vacuum fluctuations, whereas the ergoregion is seeded by the thermal Hawking quanta emanating from the horizon. This input stimulates the creation of quanta at the ergoregion, in a similar vein to ``classical'' superradiant phenomena. However, the thermal character of the input from the horizon partially inhibits the generation of entanglement by the ergoregion alone. It is only for rapidly spinning black holes ($a>0.9 M$) that the ergoregion is able to contribute significantly to the entanglement between the black hole and its exterior, completely dominating the emission in the extremal limit ($a\to 1 M$).  
 
The results of this article have significant ramifications. In the realm of black hole physics, our results emphasize that radiation enhancement by the ergoregion of a Kerr black hole is not a purely classical effect, even when the ergoregion is stimulated with classical thermal radiation. We have proven this by showing that the ergoregion is an active source of entanglement.  
 
Our findings are also important in addressing information-related questions, such as those pertaining to generalized Page curves for rotating black holes in thermal baths. For instance, by employing the techniques outlined here, one can monitor the entanglement between the escaping radiation and the interior of the rotating hole during the course of the black hole's lifetime.

Furthermore, our results can be extended to other types of ergoregions, broadening the implications beyond black hole physics. This presents an exciting opportunity for validating our predictions in analogue models, where radiation from ergoregions can actually be observed \cite{Torres_2017}.  These controllable systems are appealing because they grant access to all modes, in contrast to real black holes. Quantum fluid analogues \cite{Claude:2021rkt,Jacquet:2022vak} are  particularly promising platforms that provide a low-noise environment amenable to correlation measurements.

\begin{acknowledgements}

This article has benefited from discussion with A.~Ashtekar, M.~Jacquet and L.~Giacomelli. 
I.A.~and D.K.~are supported by the NSF grant PHY-2110273, by the RCS program of  Louisiana Boards of Regents through the grant LEQSF(2023-25)-RD-A-04, and by the Hearne Institute for Theoretical Physics. A.D. is supported by the NSF grant PHY-1903799, PHY-2206557, the Hearne Institute for Theoretical Physics, the European Regional Development Fund through the Center of Excellence TK133 “The Dark Side of the Universe”, and by the Estonian Research Council through the grant PRG356. I.A.'s research was supported in part by Perimeter Institute for Theoretical Physics. Research at Perimeter Institute is supported by the Government of Canada through the Department of Innovation, Science, and Economic Development, and by the Province of Ontario through the Ministry of Colleges and Universities. A.J.B. is supported by the DARPA Young Faculty Award (YFA) Grant No. N660012014029. 
\end{acknowledgements}


%


\appendix

\begin{widetext}

\section{Modes in Kerr geometry \label{sec:2}}

\subsection{Kerr geometry}

We collect here a few expressions which have been used in the manuscript. A detailed account of Kerr's geometry can be found, for instance, in \cite{Frolov:1998wf,Hawking:1973uf,fabbri05}.

The Kerr line element in Boyer-Lindquist coordinates takes the form 
\be
 \label{Kerr}
  ds^2=-\frac{\Delta}{\rho^2}\, (dt-a\, \sin^2\theta\, d\phi)^2+\frac{\sin^2\theta}{\rho^2}\, [(r^2+a^2)\, d\phi-a\, dt]^2+\frac{\rho^2}{\Delta}\, dr^2+\rho\, d\theta^2\, , 
\ee
where $a$ is the angular momentum of the black hole (BH) per unit mass and $\rho^2(r)=r^2+a^2\, \cos^2\theta$,  $\Delta=(r+r_+)(r-r_-)$, where  
\begin{align}
r_{\pm}=M\pm\sqrt{M^2-a^2}\, ,
\end{align}
are  the locations of the outer and inner horizons, respectively. The surface gravity and angular velocity of the horizon of a Kerr BH are
 \be
 \kappa= \frac{r_+-r_-}{2(r_+^2+a^2)}\qquad\qquad\text{and}\qquad\qquad \Omega_H= \frac{a}{r_+^2+a^2}.
 \label{eq:surfacegravity}
 \ee
For $a=M$ we have an extremal  BH, whose surface gravity vanishes, and for $a/M>1$ the Kerr line element contains a naked singularity. For this reason, we restrict to $a/M\leq 1$.  The line element \eqref{Kerr} is degenerate at $r=r_{\pm}$, but these are unphysical coordinate singularities that can be resolved by a change of coordinates.

The Kerr line element is stationary and axi-symmetric, meaning it has two independent Killing vector fields (KVF), $\partial_t$ and $\partial_\phi$.  The vector field $\partial_t$ is time-like for $r>r_{0}(\theta)=M+\sqrt{M^2-a^2\, \cos^2\theta}$, null at $r=r_{0}(\theta)$  and space-like in the interval $r_{0}(\theta)>r>r_+$. This latter region is called the ergoregion, and it plays an important role in this article since it gives rise to the phenomenon of superradiance. From the point of view of a static observer at infinity, this phenomenon is a result of the fact that $\partial_t$ is not time-like inside the ergoregion. Indeed, the KVF that becomes null at the horizon (at $r=r_+$) is the combination $\partial_t+\Omega_H\, \partial_{\phi}$. This vector field  is timelike inside the ergoregion, and it is also the generator of the event horizon. Hence, this KVF can be use to define a natural notion of frequency, $\tilde \omega$, for modes propagating near the outer horizon.

Finally, some coordinates which are useful at some points in our calculations are the tortoise coordinate $r_{\star}$, and the angular coordinate $\tilde \phi$ defined as
\be
dr_{\star}=\frac{r^2+a^2}{\Delta}\, dr \qquad\qquad\text{and}\qquad\qquad \tilde{\phi}=\phi-\Omega_H\, t.
\label{eq:tortoise}
\ee
The former has the effect of pushing the horizon to $r_\star\rightarrow-\infty$. The latter can be thought of as an angular coordinate adapted to the rotation of the spacetime at the event horizon.

\subsection{Perturbations in a Kerr spacetime: Some facts about Teukolsky equation}\label{sec:Teukolsky}

As shown by Teukolsky \cite{Teukolsky:1973ha}, perturbations of minimally coupled massless fields of spin $|s|$ in a Kerr geometry can be described by a general scalar equation,  known as Teukolsky's equation, which, quite remarkably, admits separation of variables as %
\be 
_s\phi(t,r,\phi,\theta)=\sum_{\ell,m}\int d\omega \, _sR_{\omega\ell m}(r)\, e^{-i\, \omega\, t}\, _sZ^{\omega}_{\ell m}(\theta,\phi)\, , \ee
where $s=0,\pm1/2,\pm1,\pm3/2,\pm2,...$. The angular dependence is described by spin-weighted spheroidal harmonics $_sZ^{\omega}_{\ell m}(\theta,\phi)$ (see e.g. \cite{Frolov:1998wf}). They satisfy the orthogonality relations 
\be
\int d\Omega\  _s{\bar  Z}^{\omega}_{\ell m}\, _sZ^{\omega}_{\ell'm'}=\delta_{\ell'\ell}\delta_{m m'},
\ee
 where $d\Omega$ is the standard integration measure on the sphere and we use a bar to indicate complex conjugation. The spheroidal harmonics can be defined to  satisfy
 \be
_s\bar{Z}^{\omega}_{\ell m}(\theta,\phi) =(-1)^m {}_sZ^{-\omega}_{\ell-m}(\theta,\phi).
 \ee
 The $\phi$ dependence of $_sZ^{\omega}_{\ell,m}(\theta,\phi)$ is given by a phase $e^{im\phi}$, and in the limit of zero rotation, $_sZ^{\omega}_{\ell,m}(\theta,\phi)$ reduces to the standard spin-weighted spherical harmonics (which reduce to the spherical harmonics for s=0). The spin-weighted spheroidal harmonics are solutions of a Sturm-Liouville problem. The corresponding eigenvalues $_sE_{\omega \ell m}$ are known only numerically, although analytical approximations exist in some regimes (see e.g. \cite{Berti:2005gp}). 

The radial part of Teukolsky equation can be written as follows. Define a new radial variable 
\be
_s\chi_{\omega\ell m}\equiv (r^2+a^2)^{1/2}\,  \Delta^{s/2}\, _sR_{\omega\ell m}(r).
\ee 
In terms of this variable, the radial equation takes the form
\be
\label{radialeq}
\frac{d^2}{dr_{\star}^2}\, _s\chi_{\omega\ell m}+\, _sV_{\omega \ell m }(r)\, _s\chi_{\omega\ell m}=0\, ,
\ee
where $r_{\star}$ is the  tortoise coordinate defined in \eqref{eq:tortoise}, and the effective potential $_sV_{\omega\ell m}$ has the form
\be _sV_{\omega\ell m}(r)= \frac{K^2-2is(r-M)K+ \Delta(4is\omega r-_s\lambda_{\omega\ell m})}{(r^2+a^2)^2}-G^2-\frac{dG}{dr_{\star}}\, , \ee
where 
\bea
K&=&(r^2+a^2)\, \omega-a\, m\, ,\\
G&=&\frac{r\, \Delta}{(r^2+a^2)^2}+\frac{s(r-M)}{r^2+a^2}\, ,\nonumber \\
_s\lambda_{\omega \ell m}&=&_sE_{\omega \ell m}-2 \, a \, m \, \omega+a^2\omega^2-s(s+1)\, . \nonumber \\
\eea
Although the form of this potential is complicated, it simplifies in the asymptotic regimes   
\be _sV_{\omega \ell m}(r)\sim  \left\{ \begin{matrix}\, 
   \omega^2+\frac{2is\omega}{r}  &r_{\star} \to \infty,  \\   \tilde \omega^2 -2is\tilde{\omega}\kappa-4s^2\kappa^2  &r_{\star} \to -\infty,  
     \end{matrix}\right. \,  \ee
where $\tilde \omega\equiv \omega-m\, \Omega_H$ and $\kappa$ is the surface gravity given in \eqref{eq:surfacegravity}. This implies that linearly independent solutions behave asymptotically as $r^{\pm s}e^{\pm i \, \omega\, r_{\star}}$ as $r_{\star}\to \infty$,  and $\Delta^{\pm s/2}e^{\pm i \, \tilde \omega\, r_{\star}}$ in the near horizon region $r_{\star}\to -\infty$.

\subsection{Modes for scalar waves}\label{modes}

The modes that are relevant to understand the Hawking effect are similar for all spins. Hence, the discussion of these modes and their properties can be done for scalar fields, which simplifies some technicalities which are inessential  for our main purposes. Hence, for pedagogical purposes,  we discuss first scalar modes and their properties, and discuss later  how to generalise to higher spin. Most of the material that is summarized in this section can be found in greater detail in \cite{Frolov:1998wf}. 

In the following, we write the explicit form of the modes used in the main body of the paper in the vicinity of the Cauchy hypersurface where they are defined. An important aspect of these solutions is the sign of their norm, defined from the so-called Klein-Gordon, or symplectic product. This product can be defined on any Cauchy hypersurface, and it is preserved in time for scalar fields satisfying the  Klein-Gordon equation $\Box_g\phi=0$.

It is given by 
\be \langle \varphi_1,\varphi_2\rangle =i \int_{\Sigma}d\Sigma^{\mu}(\bar \varphi_1 \partial_{\mu} \varphi_2-\varphi_2\partial_{\mu} \bar \varphi_1)\, \ee
where $d\Sigma^{\mu}$ is the oriented volume element of the Cauchy hypersurface $\Sigma$.

\subsubsection{The ``out'' modes}

These are the modes carrying Hawking radiation to future null infinity. At the Cauchy hypersurface $\Sigma_{\rm f}$, these modes have support only at $\mathcal{I}^+$, and oscillate as  $e^{- i\omega u}$. More concretely, their form in the vicinity of $\mathcal{I}^+$ is
\be \varphi^{\rm (out)}_Q \Big|_{\mathcal{I}^+}\sim  \frac{1}{\sqrt{4\pi\omega}}\, \frac{e^{-i\omega u}}{\sqrt{r^2+a^2}}\,  _0Z^{\omega}_{\ell m}(\theta,\phi)\,, \ee
where $Q=(\omega,\ell,m)$. These modes have positive norm for $\omega>0$. The ``out'' modes used in the main text are wave packets constructed from them, centered at late times $u \kappa \gg 1$, when Hawking radiation reaches $\mathcal{I}^+$. We can define normalized wave packets following Hawking's original derivation \cite{Hawking:1975vcx}. For integers $j\geq 0$ and $n$ 
\be
\varphi^{\rm (out)}_{jn \ell m}=\frac{1}{\sqrt{\epsilon}}\int^{(j+1)\epsilon}_{j\epsilon} e^{2\pi i n \omega/\epsilon}\varphi^{\rm (out)}_{\omega \ell m}\;d\omega.
\ee
These wave packets are composed of frequencies in the interval $[j\epsilon, (j+1)\epsilon]$, and are peaked at retarded time $u=2\pi n/\epsilon$ with width $2\pi/\epsilon$. Wave packets for other modes will be defined similarly. To avoid introducing new notation, we will keep using the  symbol $\varphi^{\rm (out)}_Q$ to denote these wave packets, where the label $Q$ is now a shorthand for $Q=(j,n,\ell,m)$.

\subsubsection{The ``down'' modes}

These are modes that enter the horizon at late times. In the vicinity of the  Cauchy hypersurface $\Sigma_{\rm f}$, these modes are defined to have support only at the horizon (no support at $\mathcal{I}^+$), and to oscillate as  $e^{- i \tilde \omega v}$. In more detail, they are defined as follows. First, consider the (normalized) plane waves
\be
 f^{\rm (down)}_Q\Big|_{H}=\frac{1}{\sqrt{4\pi\tilde\omega}}\, \frac{e^{-i\tilde{\omega}v}}{\sqrt{r_+^2+a^2}}\,  _0Z^{\omega}_{\ell m}(\theta,\tilde \phi)\,,
 \ee
which have a norm with a sign given by sign$(\tilde{\omega})$. From these functions, we define the down modes as
\be 
\varphi^{\rm (down)}_Q\Big|_{H}\equiv \left\{ \begin{matrix}\, 
   f^{\rm (down)}_Q  & \omega >0, \, \tilde \omega >0 \, \\  \bar f^{\rm (down)}_Q & \omega >0, \, \tilde \omega <0\, .
     \end{matrix}\right.
     \label{eq:DownModes}
\ee

These have a positive  norm for $\omega>0$. (This is due to having defined them from  the complex conjugate modes $\bar{f}^{\rm (down)}$ when $\tilde{\omega}<0$.) The ``down'' modes used in the main text, and denoted also as  $\varphi^{\rm (down)}_Q$, are normalized wave packets constructed from \eqref{eq:DownModes}, peaked at late times.

\subsubsection{The ``in'' modes}

The ``in'' modes are modes that propagate on the stationary part of the spacetime, once the BH geometry is described by the Kerr metric. They represent radiation approaching the BH from infinity. At $\Sigma_{\rm i}$, they take the form
\be
 \varphi^{\rm (in)}_Q \Big|_{\mathcal{I}^-}\sim \frac{1}{\sqrt{4\pi\omega}}\, \frac{e^{-i\omega v}}{\sqrt{r^2+a^2}}\,  _0Z^{\omega}_{\ell m}(\theta,\phi)\, .
 \label{eq:PlaneIn}
 \ee
They have a positive norm for $\omega>0$. The ``in'' modes used in the main text are normalized wave packets constructed from \eqref{eq:PlaneIn}, peaked at late times $v \kappa \gg 1$ along $\mathcal{I}^-$. 

\subsubsection{The ``up'' modes}

The ``up'' modes represent outgoing radiation trying to escape the BH. At $\Sigma_{\rm int}$, these modes have support only at the  $v=$constant part of  $\Sigma_{\rm int}$, i.e., they have no support at the horizon or at the $\mathcal{I}^-$ portions of $\Sigma_{\rm int}$. They  oscillate as  $e^{- i \tilde \omega\, u}$, and can be defined as follows. First, we define plane waves modes  as
\be 
f^{\rm (up)}_Q\Big|_{\Sigma_{\rm int}}=\frac{1}{\sqrt{4\pi\tilde{\omega}}}\, \frac{e^{-i\tilde \omega u}}{\sqrt{r^2+a^2}}\,  _0Z^{\omega}_{\ell m}(\theta,\tilde \phi)\, .
\ee
These functions have a norm with a sign given by  sign$(\tilde{\omega})$. In analogy to down modes, using these plane waves we define
\be 
\varphi^{\rm (up)}_Q\Big|_{\Sigma_{\rm int}}\equiv  \left\{ 
\begin{matrix}\, 
   f^{\rm (up)}_Q  & \omega >0, \, \tilde \omega >0  \\  \bar f^{\rm (up)}_Q & \omega >0, \, \tilde \omega <0\, , 
 \end{matrix}\right.
\ee
which have a positive norm for $\omega>0$.  
The ``up'' modes used in the main text, denoted also by $\varphi^{\rm (up)}_Q$, are normalized wave packets constructed from these modes, and supported at late retarded times $u\kappa\gg 1$ (close to the horizon). 

\subsubsection{The ``dn'' modes}

These are the partner modes of the ``up'' modes. The term ``partner'' refers to the following. In the spontaneous Hawking radiation, ``dn'' modes  are the modes that purify the thermal Hawking radiation carried in the ``up'' modes, in the sense that they are entangled with each other and with nobody else. One can find the ``dn'' modes as follows. Propagate the ``up'' modes until $\mathcal{I}^-$, and decompose the result in its positive and negative frequency contributions (frequency defined by the coordinate $v$ along $\mathcal{I}^-$)
\be \label{uppm} \varphi^{\rm (up)}_Q\Big|_{\mathcal{I}^-}= \varphi^{(+)}_Q+\varphi^{(-)}_Q\, , \ee
where we have denoted the positive and negative-frequency parts as  $\varphi^{(\pm)}_Q$, respectively. Let us denote by $\alpha_Q^2$ and $-\beta_Q^2$ the norms of $\varphi^{(\pm)}_Q$, and let us define from them positive-norm modes 
\be  \varphi^{\rm (p)}_Q\equiv \frac{\varphi^{\rm (+)}_Q}{\alpha_Q}\, {\rm and} \varphi^{\rm (d)}_Q\equiv \frac{\bar{\varphi}^{\rm (-)}_Q}{\beta_Q}\,, \ee
where $\alpha_Q$ and $\beta_Q$ can be defined real. Bringing these definitions to \eqref{uppm}, 
we obtain $\varphi^{\rm (up)}_Q=\alpha_Q\, \varphi^{\rm (p)}_Q+ \beta_Q\, \bar{\varphi}^{\rm (d)}_Q$. From this, the partner modes of ``up'' is 
\be \varphi^{\rm (dn)}_Q=\alpha_Q\, \varphi^{\rm (d)}_Q+ \beta_Q\, \bar{\varphi}^{\rm (p)}_Q\, . \ee
The form of this mode was first obtained in \cite{Wald:1975kc}, by reflecting  the mode ``up'' about $v=v_H$ at $\mathcal{I}^-$, followed by complex conjugation (the name ``dn'' is chosen because it is the reflection of ``up''). The propagation of $f^{\rm (up)}_Q$ to $\mathcal{I}^-$ produces
\be f^{\rm (up)}_Q\Big|_{\mathcal{I}^-}\sim -\frac{1}{\sqrt{4\pi\tilde\omega}}\, \frac{e^{-i \tilde{\omega}\, [v_H-\frac{1}{\kappa} \ln \kappa(v_H-v)]}}{\sqrt{r^2+a^2}}\,  _0Z^{{\omega}}_{\ell m}(\theta, \tilde\phi)\, \Theta(v_H-v)\, , \ee
where $v_H$ is the advanced time at which the horizon forms and $\Theta$ is the Heaviside function. 
From this, one finds \cite{Wald:1975kc,Frolov:1998wf}
\be f^{\rm (dn)}_Q\Big|_{\mathcal{I}^-}\sim -\frac{1}{\sqrt{4\pi\tilde\omega}}\, \frac{e^{i\tilde \omega\, [v_H-\frac{1}{\kappa}\ln \kappa(v-v_H)]}}{\sqrt{r^2+a^2}}\,  _0\bar{Z}^{{\omega}}_{\ell,m}(\theta,\tilde \phi)\, \Theta(v-v_H)\, . \ee
 From $f^{\rm (dn)}_Q$, we define the ``dn'' plane wave modes as
\be \varphi^{\rm (dn)}_Q\Big|_{\Sigma_{\rm int}}= \left\{ \begin{matrix}\, 
   f^{\rm (dn)}_Q  & \omega >0, \, \tilde \omega >0  \\  \bar f^{\rm (dn)}_Q & \omega >0, \, \tilde \omega <0\, .
     \end{matrix}\right. \, , \ee
At the Cauchy hypersurface $\Sigma_{\rm int}$, these modes have support only at the horizon, and oscillate as  $e^{- i \tilde \omega\, v}$. They have a positive norm for $\omega>0$.
From these plane wave modes, we define the ``dn'' normalized wave packets used in the main text, which are supported near $v_H$.

\subsubsection{The ``p'' and ``d'' modes}
These modes are the progenitors of the modes ``up'' and ``dn'', and can be defined from them inverting the expressions written above:
\bea \label{invsint} \varphi^{\rm (p)}_Q&=&\alpha_Q\, \varphi^{\rm (up)}_Q +\beta_Q\, \bar{\varphi}^{\rm (dn)}_Q,\ \\
\varphi^{\rm (d)}_Q&=&\alpha_Q\, \varphi^{\rm (dn)}_Q+\beta_Q\, \bar{\varphi}^{\rm (up)}_Q\, \nonumber , \eea
where $\alpha_Q=(1-e^{2\pi |\tilde \omega|/\kappa})^{-1/2}$ and $\beta_Q=e^{-\pi|\tilde \omega|/\kappa}\, (1-e^{2\pi |\tilde \omega|/\kappa})^{-1/2}$. In the main text, we have written these coefficients as $\alpha_Q=\cosh{z_H}$ and  $\beta_Q=\sinh{z_H}$, with  $z_H(\omega,m)= \tanh^{-1} e^{-|\tilde \omega|/(2\kappa)}$. 
Both modes ``p'' and ``d'' have positive unit norm for $\omega>0$. As mentioned above, both are made of a combination of positive frequency modes with respect to the coordinate $v$ in $\mathcal{I}^-$, with no contribution from negative frequency modes. This can be checked directly by computing their Fourier transform in the coordinate $v$ \cite{Wald:1975kc,Frolov:1998wf}. This automatically implies that these modes are unentangled when the field is in the vacuum state  at $\mathcal{I}^-$. In that case, the reduced quantum state describing these two modes is a pure state. This in turn proves that the state describing the pair ``up'' and ``dn'' is also  pure ---when the initial state is the vacuum--- since evolution from ``p'' and ``d'' to produce ``up'' and ``dn'' is unitary. Consequently, the ``dn'' modes are the modes that purify ``up''---i.e., these modes are the partners.

In summary, we have constructed an orthonormal wave packet mode basis of the sector of the vector space of solutions to the Klein-Gordon equation that is involved in the Hawking process, at each of the relevant Cauchy hypersurfaces: ${\Sigma_{\rm i}}$, ${\Sigma_{\rm int}}$, and ${\Sigma_{\rm f}}$. These basis elements are labeled by $Q$, with $\omega >0$, and can be collected together in (row) vectors as
$$\vec \Psi_Q^{({\rm i})}=( \varphi^{(\rm in)}_Q,\,  \varphi^{(\rm p)}_Q,\, \varphi^{(\rm d)}_Q, \bar \varphi^{(\rm in)}_Q,\,  \bar \varphi^{(\rm p)}_Q,\, \bar \varphi^{(\rm d)}_Q)\, ,$$
$$\vec \Psi_Q^{({\rm int})}=( \varphi^{(\rm in)}_Q,\,  \varphi^{(\rm up)}_Q,\, \varphi^{(\rm dn)}_Q, \bar \varphi^{(\rm in)}_Q,\,  \bar \varphi^{(\rm up)}_Q,\, \bar \varphi^{(\rm dn)}_Q)\, ,$$
$$\vec \Psi_Q^{({\rm f})}=(\varphi^{(\rm out)}_Q,\, \varphi^{(\rm down)}_Q,\, \varphi^{(\rm dn)}_Q,\,  \bar \varphi^{(\rm out)}_Q,\,  \bar \varphi^{(\rm down)}_Q,\,\bar \varphi^{(\rm dn)}_Q)\, . $$
These are called Wald bases \cite{Wald:1975kc,Frolov:1998wf}. For each $Q$, the elements of these bases are the modes that are involved in the Hawking effect. Evolution does not mix modes labeled with different $Q$. It is straightforward to extend this set of modes to have a complete basis in the space of solutions. However, since such an extension is irrelevant to describe the Hawking effect, we do not write it down.

\subsection{Evolution matrices}

In this section, we will implicitly exploit the freedom to introduce a time-independent phase to the aforementioned modes, to make the evolution matrices real.

The relation between wave packets $\vec \Psi_Q^{({\rm i})}$ and $\vec \Psi_Q^{({\rm int})}$ encodes the dynamics from ${\Sigma_{\rm i}}$ to ${\Sigma_{\rm int}}$, and is given by
\be 
\vec \Psi_Q^{({\rm i})} = \vec \Psi_Q^{({\rm int})}\cdot {\bf S}^{(\rm H)}_Q\,, \qquad\text{where}\qquad {\bf S}^{(\rm H)}_Q=\begin{pmatrix}{\bf A}_Q & {\bf B}_Q  \\ 
   {\bf B}_Q  & {\bf A}_Q 
    \end{pmatrix},
\ee
with    
\be 
{\bf A}_Q=\begin{pmatrix}1 &0 & 0   \\  0 & \cosh z_H & 0\\ 0 & 0 & \cosh z_H
\end{pmatrix}
\qquad\text{and}\qquad 
{\bf B}_Q=
\begin{pmatrix} 0 &0 & 0  \\  0 & 0 & \sinh z_H \\ 0 & \sinh z_H & 0 \end{pmatrix} \nonumber \, ,
\ee 
 where $z_{\rm H}(\omega,m)=\tanh^{-1}e^{-\pi|\tilde{\omega}|/\kappa}$. These matrices can be read off from the definitions of ``up'' and ``dn'' modes in terms of ``p'' and ``d'' modes, Eqn.~\eqref{invsint}, adding the identity transformation for the ``in'' modes. Subsequently, the evolution from $\Sigma_{\rm int}$ to $\Sigma_{\rm f}$ describes modes propagating across the ergoregion surrounding the horizon. This evolution is qualitatively different for modes with $\tilde{\omega}>0$ (non-superradiant modes, NSRM) and modes with $\tilde{\omega}<0$ (superradiant modes, SRM). For NSRM, the evolution is given by,
\be
\vec \Psi_Q^{({\rm int})} = \vec \Psi_Q^{({\rm f})} \cdot {\bf S}^{(\rm NSR)}_{Q}\;,\qquad\text{where}\qquad{\bf S}^{(\rm NSR)}_Q=\begin{pmatrix}{\bf C}_Q & 0  \\ 
 0  & {\bf C}_Q
    \end{pmatrix},
\ee 
 with
 \be  {\bf C}_Q=\begin{pmatrix} \sin \theta_Q& \cos \theta_Q & 0 \\  \cos \theta_Q & - \sin \theta_Q & 0 \\ 0& 0 & 1  \end{pmatrix},\ee 
 where $\cos^2 \theta_Q$ is the transmission coefficient of the potential barrier (the greybody factor), and is obtained by solving numerically the radial Teukolsky equation \eqref{radialeq}. $\theta_Q$ depends on $\omega$, $\ell$, $m$, and the spin of the perturbation $|s|$. Therefore, it is different for photons and gravitons.  For SRM, the evolution across the ergoregion has a completely different character and is described by 
 \be
 \vec \Psi_Q^{({\rm int})} = \vec \Psi_Q^{({\rm f})} \cdot {\bf S}^{(\rm SR)}_{Q}\;,\qquad\text{where}\qquad
 {\bf S}^{(\rm SR)}_{Q}=
 \begin{pmatrix}
	 {\bf A}^{\rm SR}_Q & {\bf B}^{\rm SR}_Q  \\ 
 	 {\bf B}^{\rm SR}_Q  &  {\bf A}^{\rm SR}_Q    
 \end{pmatrix},
\ee
with
\be
 {\bf A}^{\rm SR}_J=
 \begin{pmatrix} 
 \cosh z_{\rm erg} & 0 & 0\\
 0 &  \cosh z_{\rm erg} & 0\\
  0 & 0 & 1  \\     
  \end{pmatrix}
  \qquad\text{and}\qquad \ {\bf B}^{\rm SR}_J=
  \begin{pmatrix} 0 & \sinh z_{\rm erg} & 0  \\ 
   \sinh z_{\rm erg} & 0 & 0\\
  0 & 0 & 0
   \end{pmatrix}.
\ee
The matrix ${\bf S}^{(\rm SR)}_{Q}$ represents a {\em two-mode squeezer} between modes ``in'' and ``up'' to produce ``out'' and ``down'', and the identity evolution for the mode ``dn''.  The squeezing intensity $z_{\rm erg}$ depends on $\omega$, $\ell$, $m$, and the spin of the perturbation $|s|$, and is also obtained by solving numerically the radial Teukolsky equation \eqref{radialeq}. This implies that, for SRM, the ergoregion acts like a quantum amplifier, creating entangled pairs of particles, very much like the horizon does.

Mathematically, the distinction between the evolution matrix for SRM and NSRM arises from the fact that the Killing field describing time translations at infinity becomes spacelike within the ergoregion. For superradiant modes with $\tilde \omega<0$ but $\omega >0$, the change in the character of $\partial_t$ reverses the sign of the norm of the modes labeled as ``up" and ``down" that interact with modes ``in" and ``out". The latter two modes always have positive norm for $\omega >0$. This, in turn, implies that, for superradiant frequencies, the evolution from $\Sigma_{\rm int}$ to $\Sigma_{\rm f}$ mixes modes with different signs of their norm. At the quantum level, this translates into Bogoliubov transformations that mix creation and annihilation operators, corresponding to a quantum amplifier. This phenomenon does not occur for NSRM, as all modes with $\omega >0$ have positive norm.

This fact is also reflected in that, while both evolution matrices ${\bf S}^{(\rm SR)}_{Q}$ and ${\bf S}^{(\rm NSR)}_{Q}$ belong to the symplectic group, only ${\bf S}^{(\rm NSR)}_{Q}$ belongs to the orthogonal subgroup of the symplectic group; i.e. ${\bf S}^{(\rm NSR)}_{Q}\cdot {\bf S}^{(\rm NSR)\, \top}_{Q}=\mathbb{I}_6$. This automatically implies  that the transformation ${\bf S}^{(\rm NSR)}_{Q}$ leaves the vacuum and the particle number invariant. This is not the case for the scattering matrix of superradiant modes ${\bf S}^{(\rm SR)}_{Q}$.


We can now write  the total evolution matrix describing the evolution from $\Sigma_{\rm i}$ to $\Sigma_{\rm f}$. It is given by
\be
\vec \Psi_Q^{({\rm i})} = \vec \Psi_Q^{({\rm f})}\cdot {\bf S}^{(\rm tot)}_{Q},
\ee
where, for non-superradiant modes
\be {\bf S}^{(\rm tot)}_{Q}={\bf S}^{(\rm NSR)}_{Q}\cdot {\bf S}^{(\rm H)}_Q\, , \ee
while for SRM 
\be
 {\bf S}^{(\rm tot)}_{Q}={\bf S}^{(\rm SR)}_{Q}\cdot {\bf S}^{(\rm H)}_Q . \ee
Since evolution factorizes in decoupled $Q$ sectors, this is enough to compute the  evolution for the field. As a consistency check, one can verify that the total evolution matrices are both linear canonical transformations, meaning that they leave invariant the symplectic structure of the phase space:
\be 
\label{sympcond} 
{\bf S}^{\rm (tot)}_Q\cdot \bf{\Omega}_6\cdot {\bf S}^{\rm (tot)\top}_Q=\bm{\Omega}_6,
\qquad\text{where}\qquad
\bm{\Omega}_6=\begin{pmatrix} 0_3 & \mathbb{I}_3\\-\mathbb{I}_3& 0_3 \, .
\end{pmatrix}.
\ee
This is equivalent to saying that ${\bf S}^{\rm (tot)}_Q$ belongs to the symplectic group ${\rm Sp}(\mathbb{\mathbb{C}},6)$. Symplectic matrices have the general form 
\be
{\bf S}^{\rm (tot)}_Q=\begin{pmatrix} \bm{A} & \bm B \\\bar{\bm B}& \bar{\bm A}\end{pmatrix},
\ee where $\bm{A}$ and $\bm{B}$ are $3\times 3$ matrices and the bar denotes complex conjugation of their elements. The condition \eqref{sympcond} imposes the following constraints on the components of these matrices
\bea\bm  A\cdot \bm B^{\top}-\bm B\cdot \bm A^{\top}&=&0_3, \\
\bm A\cdot \bm A{^\dagger}-\bm B\cdot \bm B^{\dagger}&=&\mathbb{I}_3 \, .\eea
The components of $\bm{A}$ and $\bm{B}$ are usually called Bogoliubov coefficients, and these two equations are the familiar constraints satisfied by them. 

In the quantum theory, we can write the evolution in terms of creation and annihilation operators. First, let us define the column vectors of creation and annihilation operators defined from modes $\vec \Psi_Q^{({\rm i})}$, $\vec \Psi_Q^{({\rm int})}$ and $\vec \Psi_Q^{({\rm f})}$, respectively
\be \vec A_Q^{(\rm i)}=(a^{(\rm in)}_Q,\,  a^{(\rm p)}_Q,\, a^{(\rm d)}_Q, a^{(\rm in)\, \dagger}_Q,\,  a^{(\rm p)\, \dagger}_Q,\, a^{(\rm d)\, \dagger}_J)^{\top}\, , \nonumber \ee
\be \vec A_Q^{(\rm int)}=(a^{(\rm in)}_Q,\,  a^{(\rm up)}_Q,\, a^{(\rm dn)}_Q, a^{(\rm in)\, \dagger}_Q,\,  a^{(\rm up)\, \dagger}_Q,\, a^{(\rm dn)\, \dagger}_Q)^{\top}\, , \nonumber \ee
\be \vec A_Q^{(\rm f)}=(a^{(\rm out)}_Q,\,  a^{(\rm down)}_Q,\, a^{(\rm dn)}_Q, a^{(\rm out)\, \dagger}_Q,\,  a^{(\rm down)\, \dagger}_Q,\, a^{(\rm dn)\, \dagger}_Q)^{\top}\, . \nonumber \ee

The relation between ``i'' and ``f'' creation and annihilation operators, can be derived from the relation between the basis modes $\vec{\Psi}^{(\rm i)}_Q$ and $\vec{\Psi}^{(\rm f)}_Q$ using the following argument. The field operator $\phi(x)$ can be expanded in any of these two sets of modes

\be \label{phi} \phi(x)=\sum_Q \vec{\Psi}^{(\rm i)}_Q(x)\cdot \vec{A}^{(\rm i)}_Q=\sum_Q \vec{\Psi}^{(\rm f)}_Q(x)\cdot \vec{A}^{(\rm f)}_Q\, .\ee

If we start from the ``i'' representation, $\phi(x)=\sum_Q \vec{\Psi}^{(\rm i)}_Q(x)\cdot \vec{A}^{(\rm i)}_Q$ and replace $\vec{\Psi}^{(\rm i)}_Q$ = $\vec{\Psi}^{(\rm f)}_Q(x)\cdot {\bf S}^{(\rm tot)}_{Q}$, we find 
$$\phi(x)=\sum_Q \vec{\Psi}^{(\rm f)}_Q(x)\cdot {\bf S}^{(\rm tot)}_{Q}\cdot \vec{A}^{(\rm i)}_Q\, .$$
Comparing this result with the last part of \eqref{phi}, we conclude
 $$\vec A_Q^{(\rm f)}= {\bf S}^{(\rm tot)}_Q\cdot   \vec A_Q^{(\rm i)}.$$
 The commutation relation satisfied by $\vec A_Q^{(\rm i)}$ can be compactly written in vector notation as
\be [\vec A_Q^{(\rm i)},\vec A_Q^{(\rm i)}]=\bm{\Omega}_6 \, . \ee
The fact that $ {\bf S}^{(\rm tot)}_Q$ is a symplectic transformation, guarantees that  the operators $\vec A_Q^{(\rm f)}$ satisfy the same commutation relations:
\be [\vec A_Q^{(\rm f)},\vec A_Q^{(\rm f)}]=  [ {\bf S}^{(\rm tot)}_J\cdot \vec A_Q^{(\rm i)},{\bf S}^{(\rm tot)}_Q\cdot  \vec A_Q^{(\rm i)}]={\bf S}^{(\rm tot)}_Q\cdot \underbrace{[\vec A_Q^{(\rm i)},\vec A_Q^{(\rm i)}]}_{=\bm{\Omega}_6} \cdot {\bf S}^{(\rm tot)\, \top}_Q=\bm{\Omega}_6\, \ee

\subsection{Electromagnetic and gravitational waves}

The strategy to construct the Wald basis at each of the Cauchy hypersurfaces for massless higher spin fields is conceptually similar to the scalar case. The technical details are, however,  more involved, in part due to the presence of polarization tensors and gauge redundancies. Fortunately, this problem was solved already in the seventies and early eighties in \cite{PhysRevD.11.2042,Wald:1978vm,Candelas:1981zv,Galtsov:1982hwm,Galtsov:1986rzw,Futterman:1988ni}.  A complete summary, containing many useful details, can be found in Appendix G of \cite{Frolov:1998wf}.

Electromagnetic and gravitational perturbations are also described by the Teukolsky equation for $|s|=1$ and $|s|=2$, respectively. A basis of solutions to this equation can again be labeled by $\omega,\ell, m, s$, where the sign of $s$ denotes the two possible polarization states (helicity). These modes are described by covariant vectors
 $a_{\mu}(\omega,\ell, m, s, x)$ for electromagnetic fields, and symmetric covariant tensors $h_{\mu\nu}(\omega,\ell, m, s, x)$ for gravitational linear perturbations. In an analogous way as done in the previous section, one defines modes  ``out'', ``in", ``down'', ``up'', ``dn'', ``p'' and ``d''. In Kerr geometry, explicit expressions for these modes are known. These expressions are, however, complicated and not very illuminating. Furthermore, they are not necessary for our goals. Rather, the important aspect of these solutions is that their $u$, $v$ and $\phi$ dependence on the three Cauchy hypersurfaces $\Sigma_{\rm f}$, $\Sigma_{\rm int}$ and $\Sigma_{\rm i}$ is the same as for the scalar case. In particular, the evolution matrix describing the propagation from $\Sigma_{\rm i}$ to $\Sigma_{\rm int}$, is precisely the same for electromagnetic and gravitational perturbations as for scalar fields. Heuristically, this can be understood as a consequence of the exponentially large  blueshift near the event horizon, which makes conformal invariance to emerge in the near horizon region \cite{Agullo:2010hi}. The evolution from  $\Sigma_{\rm int}$ to $\Sigma_{\rm f}$, in contrast, depends on the spin of the field, and is dictated by the Teukolsky equation. 

Following Teukolsky and Press \cite{Teukolsky:1974yv}, as well as Page's calculations \cite{Page:1976ki}, we have obtained the evolution matrix by solving numerically the radial Teukolsky equation for spin $|s|=1$ and $|s|=2$ for different values of the black hole spin  $a$. As explained by Page, the dependence of the matrix elements on the BH parameters is through the dimensionless quantities $M\omega$ and $a^\star=a/M$. Also, for massless particles, their dependence on the particle species, spin, and helicity is only through $|s|$ and the number of polarizations, which is 2 for both photons and gravitons. Aside from that, the matrix elements also  depend on the angular momentum labels $\ell$ and $m$ of the modes. We have solved Teukolsky equation and obtained the evolution matrix for a range of frequencies $\omega\in[0,\omega_{\rm max}]$ and within each frequency for $\ell\in[|s|,\ell_{\rm max}]\subset \mathbb{N}$, and for $m\in[-\ell,\ell]\subset \mathbb{Z}$. We have chosen $\omega_{\rm max}$ and $\ell_{\rm max}$ in such a way that the contribution of larger frequencies or $\ell$ to the energy, angular momentum and entanglement radiated away by the BH is negligible 
 within the precision of our numerical calculation. Such maximum values depend on the spin parameter $a^\star$. In particular, we have make sure that the upper value of the frequency is high enough as to explore the SR to NSR transition for each of the modes that are solved.

\subsection{Our numerical methods}\label{num}

The evolution matrix that relates modes  ``up'' and ``in'' with modes ``out'' and ``down'' (corresponding to evolution from the Cauchy hypersurfaces $\Sigma_{\rm int}$ to $\Sigma_{\rm out}$) is not analytically solvable. We briefly summarize in this subsection the numerical methods we use to obtain the evolution matrix and provide references to earlier work upon which we have built our calculations. 

The goal is to solve the radial Teukolsky equation, which is solved numerically with purely ingoing boundary conditions at the horizon. This equation depends on  eigenvalues of the spin-weigthed spheroidal harmonics, which are obtained through the small-c expansion method in section II.C of \cite{Berti:2005gp}, taking up to the 5th order term in an expansion in $\omega a$.
In our calculations, we have checked that the first term being neglected in the expansion --- the 6th order term --- would lead to negligible corrections to our results for total entanglement radiated even in the worst case scenario when the black hole spin approaches $a=1$. The sources of error in our calculations and their estimated  contribution are provided below.

To solve the radial equation, we use standard integration algorithms included in the software Mathematica through the in-built function NDSolve, with specifications AccuracyGoal$=10$, PrecisionGoal$=10$ and MaxStepSize$=0.1$. This function uses an adaptative method, which might make use of different numerical integrator algorithms in each interval, based on the problem to solve.

In order to avoid that numerical errors in the initial conditions excite unphysical outgoing modes at the horizon, we follow Teukolsky and Press (section VII \cite{Teukolsky:1974yv}). Thus, we use a modified radial coordinate which vanishes at the horizon, and is defined by $x\equiv  (r-r_+)/r_+$, and a modified radial function that differs from the original variable $_sR_{\omega,\ell,m}$  in a polynomial in the variable $x$ of order $|s|+1$, where $|s|$ is the spin of the field. This spin-dependent polynomial is computed by requiring that the ``up'' (outgoing) modes approach zero asymptotically faster than the ``down'' (ingoing) modes when approaching the horizon. This guarantees the numerical stability when integrating outwards for ingoing boundary conditions, which are provided near the horizon, at $x=10^{-15}$, following \cite{Teukolsky:1974yv}. In this way we obtain the sought modes of the system, from which the greybody factors can be computed by requiring conservation of the flux(particle numbers and entanglement remain independent of the phases of the transmission and reflection coefficients, requiring us only to compute their amplitudes). This calculation was
carried out in \cite{Teukolsky:1974yv} as outlined in section VII, leading to formulas A.17 of that work, which relate the greybody factors and the amplitudes of the modes far from the black hole. We extract these numbers at $x=700$ (i.e. $r=701r_+$). Although we have created our own code \footnote{We thank Adrian del Rio for useful guidance on this calculation in the early stages of this project.}, publicly available resources exist to do this calculation \cite{CodesSuperradiace}. 

We have solved Teukolsky equation and obtained the evolution matrix for a range of frequencies  $\omega\in[0.01M^{-1},\omega^{\ell}_{\rm max}]$, where $\omega^{\ell}_{\rm max}$ is an $\ell$-dependent cutoff frequency, and $\ell$'s within the interval $\ell\in[|s|,\ell_{\rm max}]\subset \mathbb{Z}$. 

We have chosen $\ell_{\rm max}=5$ for photons and $\ell_{\rm max}=6$ for gravitons. The cutoff frequencies   $\omega_{\max}^{\ell}$ used in our calculations are shown in Table \ref{tab:Powerls}. The contribution above these cutoffs as well as the error in the angular eigenvalues are the main sources of error in our calculations. For these choices of $\ell_{\rm max}$ and $\omega^\ell_{\rm max}$, the error introduced in the determination of the angular eigenvalues is at the percent level at most. Concretely, for $a=0.99$, we have checked that the correction of the first term ignored in the small-c expansion, namely the 6th order term, corrects the eigenvalues below the $0.25\%$. Furthermore, this error decreases for lower values of $a$ and $\omega$. Since errors are smaller for the low frequencies ---which carry most of the energy radiated--- the actual error in the determination of the angular eigenvalues weakly impacts integrated quantities. In particular, we have verified that the relative errors in the total power emitted in each angular multipole by frequencies above $\omega_{\rm max}^\ell$ are below $10^{-5}\%$ with respect to the total energy radiated by modes with that value of $\ell$.

To quantify the total error  in the integrated spectrum we would have to take into account the contribution of the missing modes. In particular, the missing higher multipoles $\ell>\ell_{\rm max}$ will be the main source of errors, since $\omega>\omega_{\rm max}^\ell$ contribute well below the $0.17\%$ contribution of the $\ell=5$ photon modes to total emission (the lowest of the considered contributions). 
 This is difficult to compute exactly, but taking into account the exponential decay of the emitted power with the value of the angular multipole, both for gravitons and photons, we estimate that the corrections to total entanglement radiated are below the percent level.\\

\begin{table}[h!]
\setlength{\extrarowheight}{5pt}
    \centering
    \begin{tabular}{|c|c|c|c|c|c|}
    \hline
           Photons & $\ell=1$ & $\ell=2$ & $\ell=3$ & $\ell=4$ & $\ell=5$ \\
          \hline
            $\omega_{\rm max}^\ell$ & 1 & 1.5 & 2 & 2.3 & 2.7\\
          \hline
         $dP/dt(\omega<\omega_{\rm max}^{\ell})$& $9.2\times10^{-4}$ & $6.4\times10^{-4}$ & $3.3\times10^{-4}$ & $1.5\times10^{-4}$ & $6.4\times10^{-5}$\\
         \hline
         \% of total $dP/dt$ & $2.44\%$ & $1.71\%$ & $0.87\%$ & $0.40\%$ & $0.17\%$\\
    \hline
    \end{tabular}
        \vspace{.5cm}
    
    \begin{tabular}{|c|c|c|c|c|c|}
    \hline
           Gravitons &  $\ell=2$ & $\ell=3$ & $\ell=4$ & $\ell=5$ & $\ell=6$\\
          \hline
           $\omega_{\rm max}^\ell$ & 1.2 & 1.5 & 2 & 2.4 & 2.8\\
          \hline
         $dP/dt(\omega<\omega_{\rm max}^{\ell})$&  $2.59\times10^{-2}$ & $6.88\times10^{-3}$ & $1.96\times10^{-3}$ & $5.95\times10^{-4}$ & $1.90\times10^{-4}$\\
         \hline
         \% of total $dP/dt$ & $68.84\%$ & $18.28\%$ & $5.21\%$ & $1.58\%$ & $0.50\%$\\
    \hline
    \end{tabular}
    \caption{These tables contain information about the emission in each multipole for $a=0.99$. 
    The second row in each table shows the $\ell$-dependent cutoff frequency $\omega^\ell_{\rm max}$, in units of $M^{-1}$, for photons and gravitons. The third row provides the power radiated in modes lying in the interval $\omega\in[0.01,\omega_{\rm max}^\ell]$. The fourth row shows the contribution to the total power radiated in each multipole $\ell$ ---summed for all allowed $m$'s and for photons and gravitons--- normalized to a 100\% (the large relative contribution of gravitons is due to the large value of $a$ used in this table).  Note that the contribution to the total power decays exponentially with $\ell$, so the contribution of higher $\ell$'s is negligible. For smaller black hole spin $a$ the main contribution to the power radiated is further shifted toward small $\ell$.}
    \label{tab:Powerls}
\end{table}
\section{Gaussian states and their evolution \label{sec:3}}\label{Gauss}

We summarize some general properties of Gaussian bosonic systems.

\subsection{Gaussian states}
Let us consider a dynamical system containing $N$ classical degrees of freedom. Quantum mechanically, the system is described by $N$ pairs of canonically conjugate operators $(x_I, p_I)$ with $I=1,...,N$ or, equivalently, by $N$ pairs of creation and annihilation operators $(a_I,a^{\dagger}_I)$. The creation and annihilation operators are related to the canonical variables $x$ and $p$ by 
\be
\begin{pmatrix}
a_I\\
a^\dagger_I
\end{pmatrix}
=\frac{1}{\sqrt{2}}
\begin{pmatrix}
1 & i\\
1 & -i
\end{pmatrix}
\begin{pmatrix}
x_I\\
p_I
\end{pmatrix}
\qquad\text{and}\qquad
\begin{pmatrix}
x_I\\
p_I
\end{pmatrix}
=\frac{1}{\sqrt{2}}
\begin{pmatrix}
1 & 1\\
-i & i
\end{pmatrix}
\begin{pmatrix}
a_I\\
a^\dagger_I
\end{pmatrix} .
\ee
The discussion below can be made using either sets of variables. We will use creation and annihilation operators. Let us define a vector $\vec A$ by putting together all creation and annihilation operators in the following order
\be \vec A=(a_1,\cdots,a_N,a_1^{\dagger},\cdots a_N^{\dagger})^{\top}\, , \ee
where the transposition symbol indicates we define this as a column vector (in the literature,  other orderings can be found, such as $(a_1,a^\dagger_1,...,a_N,a^\dagger_N)^{\top}$. It is straightforward to translate expressions written below from one choice of order to another.) 

 Gaussian states $\rho$, either pure or mixed, are states for which the quantum moments $\langle {A}^{i_1}\cdots {A}^{i_n}\rangle$ satisfy the same relations as the statistical moments of a Gaussian multi-variable probability distribution. In particular,  all moments can be determined from the first and second moments. In other words, the first and the second moments contain complete information about the state, and they can be used as to describe a quantum Gaussian state. The first moments can be encoded in a  $2N$ dimensional vector  ${\vec \mu}= \langle \vec A \rangle$, while the second moments are conveniently encoded in the so-called covariance matrix, defined as $\sigma^{ij}=\langle \{ A^i-\mu^i,A^j-\mu^j) \rangle$, where $i,j=1,...,2N$ and the curly bracket denotes the anti-commutator. We use the symbol $\langle \cdot\, ,\, \cdot \rangle$ to indicate expectation value,  for both pure and mixed states.  In the definition of the covariance matrix, the subtraction of  $\vec{\mu}$ guarantees that the result is independent of the first moments. The anti-commutator selects the symmetric part of the second moments, since the anti-symmetric part, i.e., the commutator, is state-independent and completely determined from the canonical commutation relations. 
 
Many properties of a Gaussian state can be easily extracted only from its covariance matrix ${\sigma}$. As an example, one such property is the purity $P$ of the state, which is completely independent of the first moments. For a Gaussian state, it is obtained from its covariance matrix as $P(\sigma)=1/\sqrt{{\rm{det}\sigma}}$ (it is one for pure states and smaller than one for mixed states). Another quantity that is completely determined only by the covariance matrix is the Logarithmic Negativity, which serves as a faithful entanglement quantifier in some cases (see below).

For classical systems, if the Hamiltonian dictating the dynamics is a quadratic function of the canonical variables, the evolution is a linear canonical transformation $\bm{S}$. Linear evolution preserves the Gaussianity of states. If $(\vec{\mu}^{\rm (in)},{\sigma}^{\rm (in)})$ describes a quantum Gaussian state, evolution under a quadratic Hamiltonian maps it to another Gaussian state, described by 
\begin{align} 
\vec{\mu}^{\rm(out)}&=\bm{S}\cdot\vec{\mu}^{\rm(in)},\label{eq:out_mu_general}\\
{\sigma}^{\rm(out)}&=\bm{S}\cdot{\sigma}^{\rm(in)}\cdot\bm{S}^\top,\label{eq:out_sigma_general}
\end{align}
where $\bm{S}$ agrees with the evolution matrix describing the classical dynamics.
 
\subsubsection{Entropy of  Gaussian states}  

The von Neumann entropy of a Gaussian state  $({\vec \mu},{\sigma})$ describing a system with $N$ degrees of freedom can be easily computed from the $N$ symplectic eigenvalues of  ${\sigma}$. These are $N$ real numbers, denoted by $\nu_I$, with $I=1,..., N$,  defined as  the 
absolute value of the  eigenvalues of the matrix $\sigma\cdot\bm{\Omega}^{-1}$
where $\bm{\Omega}^{-1}$ is the inverse of the symplectic form.  [Note there are technically $2N$ eigenvalues, since 
 $\sigma\cdot\bm{\Omega}^{-1}$ is a $2N\times 2N$ matrix, but the eigenvalues come in  pairs $\pm \nu_I$.] 
In terms of $\nu_I$, the von Neuman entropy of a Gaussian state reads 
\bea 
S[\sigma]=\sum_I^N&\Big[& \left( \frac{\nu_I+1}{2}\right) \ln\left( \frac{\nu_I+1}{2}\right)-\left( \frac{\nu_I-1}{2}\right) \ln\left( \frac{\nu_I-1}{2}\right)\Big] . 
\eea

As an example, for a Gaussian state with $\vec{\mu}=\vec{0}$ and $\sigma = \bigoplus_I (1+2\bar{n}_I)\bar{I}_2$---which holds for outgoing radiation of an evaporating black hole (even with thermal inputs)---we have $\nu_I=(1+2\bar{n}_I)$
, and we can write the entropy as $S[\sigma]=\sum_{I=1}^N (\bar{n}_I+1)\ln(\bar{n}_I+1)-\bar{n}_I\ln(\bar{n}_I)$, which is just the standard ``thermal entropy''. 

\subsubsection{Entanglement  in Gaussian states}

Consider a partition of the system of $N$ modes into two subsystems $A$ and $B$, defined by two disjoint subsets of pairs of creation and annihilation operators $(a_I, a^{\dagger}_I)$. If the total state of the system is Gaussian $(\vec \mu_{AB},\sigma_{AB})$, the reduced states describing each subsystem are also Gaussian. The vector of first moments and the covariance matrices describing each of the Gaussian subsystems can be obtained from $(\vec \mu_{AB},\sigma_{AB})$ by simply restricting to the components associated with each subsystem.

If the total state $(\vec \mu_{AB},\sigma_{AB})$ is pure, then the von Neumann entropies of the reduce systems satisfy $S[\sigma_A]=S[\sigma_B]$. Moreover, the entropy in this case quantifies the entanglement between $A$ and $B$---i.e., the von Neumann entropy is the entanglement entropy. On the other hand, if the total state is mixed, the von Neuman entropy of the subsystems can be different, and the entropy does not quantify entanglement. 

A convenient entanglement measure, useful for pure and mixed states alike, is the Logarithmic Negativity, ${\rm LN}$. It is defined as
\be {\rm LN}( \rho)=\log_2 || \rho^{\top_B}||_1\, \label{eq:defLN2},\ee
where $\rho$ is the density matrix of the system, $\rho^{\top_B}$ is its partial transpose with respect to subsystem $B$ (the same result is obtained if we partial transpose in subsystem A instead), and $||\cdot||_1$ is the trace norm. The Logarithmic Negativity quantifies the violation of the  PPT criterion (Positivity of Partial Transpose) for quantum states \cite{peres96,simon2000criterion,plenio05}. This criterion is satisfied by all separable states and, as a consequence, its violation signals the existence of entanglement. For a Gaussian state made of two Gaussian subsystems A and B, Eqn.~\eqref{eq:defLN2} can be rewritten in terms of the covariance matrix ${\sigma}_{AB}$ as \cite{serafini17QCV} 
\be \label{LN} {\rm LN}[{\sigma_{AB}}]=\sum_I {\rm Max}[0, -\log_2 \tilde \nu_I]\, , \ee
where $\tilde \nu_I$ are  the  symplectic eigenvalues of the {\em partially transposed} state. The eigenvalues $\tilde \nu_I$ are computed as follows.

Let $N_A$ and $N_B$ denote the number of modes within subsystems A and B, respectively. Hence, $\sigma_{AB}$ is a $2\times(N_A+N_B)$ matrix. It is not difficult to notice that the covariance matrix of $ \rho^{\top_B}$ can be obtained from the covariance matrix of $\rho$ by interchanging creation and annihilation operators in each mode within subsystem B. Furthermore, this interchange can be operationally implemented via
%
\be \tilde \sigma_{AB}= {\bm T}\sigma_{AB}{\bm T}\, \ee
where 
\be {\bm T}=\begin{pmatrix} \mathbb{I}_{N_A} & 0_{N_AN_B}& 0_{N_A}& 0_{N_AN_B}\\ 0_{N_BNA} & 0_{N_B}& 0_{N_BN_A}& \mathbb{I}_{N_B}\\ 0_{N_A} & 0_{N_AN_B}& \mathbb{I}_{N_A} & 0_{N_AN_B}\\ 0_{N_BN_A} & \mathbb{I}_{N_B} & 0_{N_BN_A}& 0_{N_B} \end{pmatrix}\, . \ee
In this expression, $\mathbb{I}_{N}$ is the $N$-dimensional identity matrix, $0_{N}$ is the $N$-dimensional zero matrix, and $0_{N_AN_B}$ is a $N_A\times N_B$ matrix with all components equal zero.

With this, the $N_A+N_B$ positive real numbers $\tilde \nu_I$ featuring in the expression for  ${\rm LN}[{\sigma_{AB}}]$ are the absolute values of the eigenvalues of the matrix $\tilde \sigma_{AB} \cdot \bm{\Omega}^{-1}$.

With this information, the calculation of Logarithmic Negativity generated in Hawking's process is performed as follows: Starting from the covariance matrix of the initial state $\sigma_{\rm in}$ for each wave packet mode $Q=(j, n,\ell, m, s)$, we obtain the ``out" covariance matrix as $\sigma_{\rm out}=\bm{S}_{Q}^{\rm (tot)}\cdot \sigma_{\rm in}\cdot \bm{S}^{\rm (tot)\, \top}_{Q}$, where $\bm{S}_Q^{\rm (tot)}$ is the scattering matrix containing the scattering coefficients for the mode $Q$. From $\sigma_{\rm out}$, we compute the reduced covariance matrix of the set of modes we are interested in (e.g., mode out and dn), then compute  the partial transpose with respect to one of the subsystems (e.g., mode out), evaluate its symplectic eigenvalues, and substitute them into \eqref{LN}.

For a Gaussian state, non-vanishing LN is a sufficient condition for entanglement, but it is not necessary in general. Therefore, there exist  Gaussian states which are entangled but have vanishing LN. However, if either of the Gaussian subsystems is made of a single degree of freedom, (e.g. $N_A=1$),  ${\rm LN}>0$ is also a necessary condition for the presence of entanglement, regardless of the size of the other subsystem. LN is also an entanglement monotone and, therefore, can be used to quantify entanglement (see \cite{serafini17QCV} for further details).  In the main body of this article we have used LN only for Gaussian states in which one of the subsystems has a single mode, so that it is a legitimate entanglement quantifier for our purposes. As a last remark, Logarithmic Negativity  has an operational meaning for Gaussian states: it is the exact cost (measured in Bell pairs or entangled bits, ebits) that is required to prepare or simulate the quantum state under consideration \cite{wilde2020ent_cost, wilde2020alpha_ln}. It is also an upper bound for distillable entanglement. 

\end{widetext}

\end{document}